\documentclass[twocolumn,aps,showpacs,floatfix,superscriptaddress]{revtex4-1}

\usepackage{graphicx}
\usepackage{rotating}
\usepackage{amsmath}
\usepackage{bbm}
\usepackage{subfigure}
\usepackage{color}

\renewcommand{\vec}[1]{{\bf #1}}
\renewcommand{\v}[1]{{\bf #1}}

\newcommand{\be}{\begin{equation}}
\newcommand{\ee}{\end{equation}}
\newcommand{\bea}{\begin{eqnarray}}
\newcommand{\eea}{\end{eqnarray}}

\graphicspath{
  {figures/pdf/}
  {figures/eps/}
}

\begin{document}
\title{Non-linear phenomena in time-dependent density-functional theory: What Rabi physics can teach us}

\author{J.I. Fuks}
\affiliation{Nano-Bio Spectroscopy group and ETSF Scientific Development Center, Dpto.~F\'isica de Materiales, Universidad del Pa\'is Vasco, Centro de
F\'isica de Materiales CSIC-UPV/EHU-MPC and DIPC, Av.~Tolosa 72, E-20018 San 
Sebasti\'an, Spain}

\author{N. Helbig}
\affiliation{Nano-Bio Spectroscopy group and ETSF Scientific Development Center, 
Dpto.~F\'isica de Materiales, Universidad del Pa\'is Vasco, Centro de
F\'isica de Materiales CSIC-UPV/EHU-MPC and DIPC, Av.~Tolosa 72, E-20018 San 
Sebasti\'an, Spain}
\author{I.V. Tokatly}
\affiliation{Nano-Bio Spectroscopy group and ETSF Scientific Development Center, 
Dpto.~F\'isica de Materiales, Universidad del Pa\'is Vasco, Centro de
F\'isica de Materiales CSIC-UPV/EHU-MPC and DIPC, Av.~Tolosa 72, E-20018 San 
Sebasti\'an, Spain}
\affiliation{IKERBASQUE, Basque Foundation for Science, E-48011 Bilbao, Spain}
\author{A. Rubio}
\affiliation{Nano-Bio Spectroscopy group and ETSF Scientific Development Center, 
Dpto.~F\'isica de Materiales, Universidad del Pa\'is Vasco, Centro de
F\'isica de Materiales CSIC-UPV/EHU-MPC and DIPC, Av.~Tolosa 72, E-20018 San 
Sebasti\'an, Spain}
\affiliation{Fritz-Haber-Institut der Max-Planck-Gesellschaft, Faradayweg 4-6,
D-14195 Berlin, Germany}

\begin{abstract}
Through the exact solution of a two-electron singlet system interacting with a
monochromatic laser we prove that all adiabatic density functionals within
time-dependent density-functional theory are not able to discern between
resonant and non-resonant (detuned) Rabi oscillations. This is rationalized in
terms of a fictitious dynamical exchange-correlation (xc) detuning of the
resonance while the laser is acting. The non-linear dynamics of the Kohn-Sham 
system shows the characteristic features of detuned Rabi oscillations even if
the exact resonant frequency is used. We identify the source of this error in a
contribution from the xc-functional to the set of non-linear equations that 
describes  the electron dynamics in an effective two-level system.  The
constraint of preventing the detuning introduces a new strong condition to be
satisfied by approximate xc-functionals.
\end{abstract}

\date{\today}

\maketitle

\section{Introduction}\label{sec:intro}

Despite of the success of linear-response schemes to describe excitations of
many electron systems, many physical processes stemming from the interaction of
light with matter are non-linear in nature (i.e.\ photo-reactivity, population
quantum control, etc). A prototype effect, ubiquitous in many fields of physics,
are Rabi oscillations between the ground and an excited state when a
monochromatic laser with a  frequency close to the resonance is applied. The
dipole moment of the system then shows fast oscillations with the frequency of
the applied laser and an envelope that oscillates with the frequency attributed
to the transition from the ground to the excited state (Rabi frequency). It is one of the few
analytically solvable cases of non-linear electron dynamics where the population
of states changes dramatically in time. Thus, Rabi dynamics provides a firm
ground to derive fundamental properties of the exact exchange-correlation  (xc)
functional of time-dependent density-functional theory (TDDFT)
\cite{RG1984,TDDFT2006}. This is one of the scopes of the present paper
together with providing a sound description of laser induced population
processes within TDDFT. We remark that it was recently concluded that Rabi
oscillations cannot be described within a finite-level model in TDDFT
\cite{BauerRugg}. The authors observed that the dipole moment in an adiabatic
exact exchange (EXX) calculation shows the characteristic Rabi oscillations
while the  density does not show a transition to the excited state.

Here, we show that for all adiabatic functionals the observed oscillations in
the dipole moment can indeed be described as Rabi oscillations, however, with a
non-resonant laser frequency. The employed adiabatic approximation results in a
detuning, i.e.\ the system is driven out of resonance by the change in the
Kohn-Sham (KS) potential which is due to the change in the density.  The failure
of adiabatic density functionals to correctly describe Rabi oscillations has
some  resemblance to well-known problems of describing double and charge
transfer excitations \cite{MZCB2004,DH2004}  and ionization processes
\cite{PG1999} that were all traced back to the lack of memory in the
functional.  In the present work we use the example of Rabi oscillations to
identify the physics behind the failure of adiabatic  constructions to describe
resonant non-linear dynamics. Our results show that the problems are not limited
to adiabatic  TDDFT but affect any mean-field theory.

The paper is organized as follows. In Section \ref{sec:rabi} we derive the
analytic solution for Rabi oscillations in a many-body interacting system and
apply this solution to compute the dipole moment and the populations of an
exactly solvable one-dimensional model. In Section \ref{sec:rabiks} we map the 
interacting problem onto a non-interacting Kohn-Sham one and carry out numerical
time propagations using two different adiabatic approximations, ALDA and EXX. We
also derive analytic expressions for the dipole moment and the populations in
the KS system within the adiabatic EXX approximation and discuss and quantify
the appearing dynamical detuning. We conclude our paper with a summary and outlook in
Section \ref{sec:concl}.


\section{Rabi dynamics in interacting many-body systems}\label{sec:rabi}
The Hamiltonian for an interacting non-relativistic $N$-electron system coupled 
to a laser field of the form ${\cal E}(t)= {\cal E}_0 \sin(\omega t)$ in dipole 
approximation is given by
\begin{equation}
\hat{H}= \hat{T} + \hat{V}_{ext} + \hat{V}_{ee}  + \sum_{j=1}^N \vec{r}_j \vec{{\cal E}}(t)
\label{eq:LinHgeneral}
\end{equation}
with $\hat{T}$ being the kinetic energy, $\hat{V}_{ext}$ describing the external 
potential, and $\hat{V}_{ee}$ being the electron-electron interaction (atomic 
units $m=e=\hbar=1$ are used throughout the paper). We denote the 
eigenstates and eigenvalues of Hamiltonian (\ref{eq:LinHgeneral}) with $\psi_k$ and $\epsilon_k$, respectively. In order for Rabi's solution to be valid (for a detailed discussion see e.g.\ Ref.~\onlinecite{QMTD}), 
the system under study needs to be an effective two-level system reducing the solution space to $\psi_g$ (ground state) and $\psi_e$ (dipole allowed excited state) with eigenenergies $\epsilon_g$ and $\epsilon_e$. 
The system then has a resonance at $\Delta=\epsilon_e-\epsilon_g$. 
For a generic situation the two-level approximation should be valid if the laser frequency $\omega$ is close to the resonance, and the amplitude of the driving field is not too strong to disturb the rest of the spectrum. 
These conditions can be formalized as follows
\begin{equation}
 {\delta}<< {\Delta}, \quad  \Omega_0 << \omega,
\label{eq:cond}
\end{equation}
where $\delta= \omega -\Delta$ is the detuning from the two-level 
resonance and $\Omega_0=d_{eg}{\cal E}_0$ is the resonant Rabi frequency,
with $d_{eg}=\langle \psi_g | \sum_j\hat{\v r}_j | \psi_e \rangle$ being the 
dipole transition matrix element. 

\begin{figure}
\includegraphics[width=0.47\textwidth]{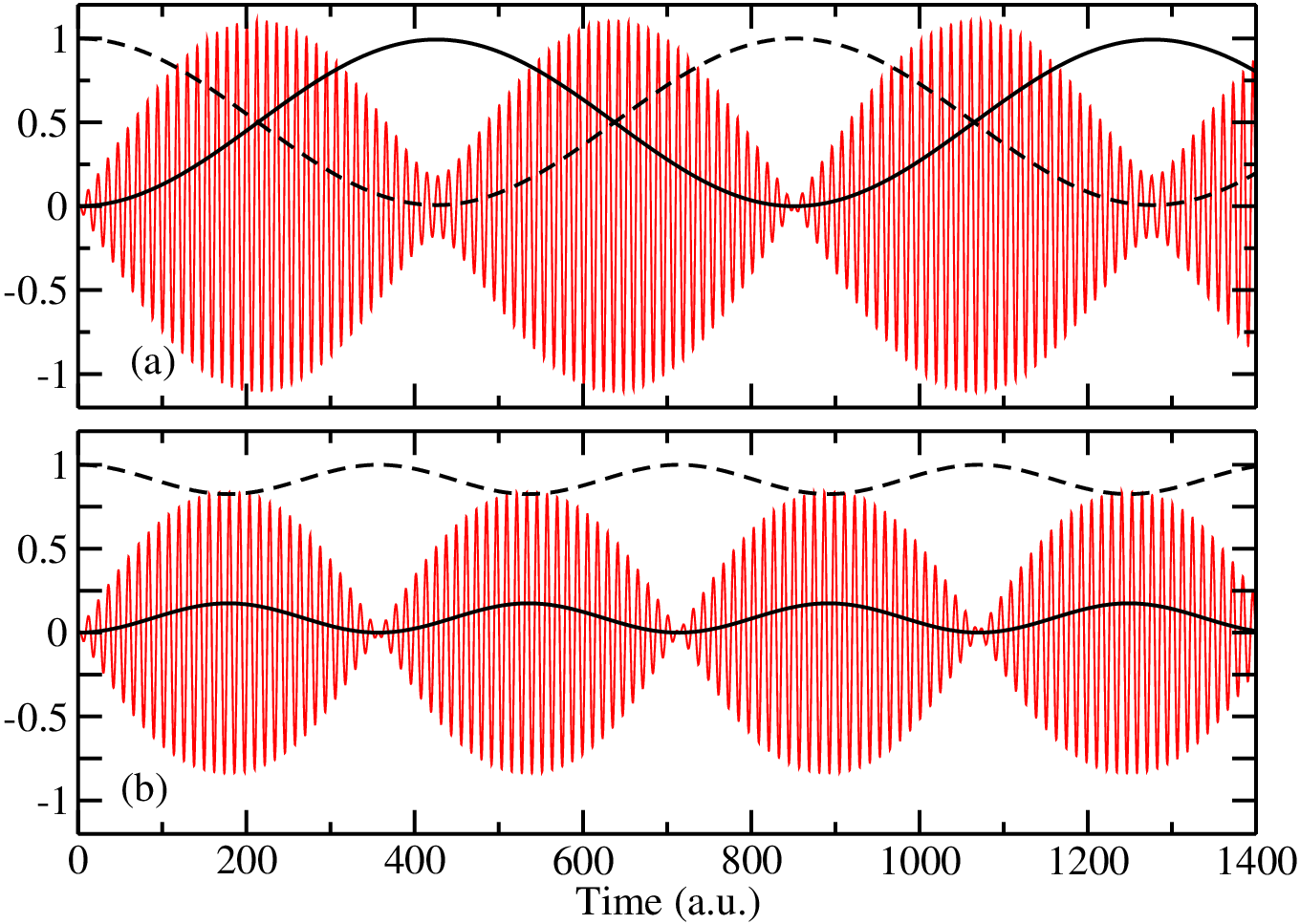}
\caption{\label{fig:dipolen1Lin}  Dipole moment (red) and populations 
$n_e$ (solid black line) and $n_g$ (dashed black line) from the analytic solution of 
(\ref{eq:proj1}) using ${\cal E}_0= 0.0125 \omega$ for detuning $\delta=0.08 \Omega_0$ ($0.0006$ Ha) (a) 
 and $\delta=2.2 \Omega_0$ ($0.016$ Ha) (b).}
\end{figure}

For an effective two-level system the time-dependent many-body state 
$|\Psi(t)\rangle$ can be written as a linear combination of both ground and 
excited states, i.e.
\be \label{eq:2levellin}
 |\psi(t)\rangle = a_g(t) |\psi_g\rangle  + a_e(t) |\psi_e\rangle
\ee
with $|a_g(t)|^2=n_g (t)$ and $|a_e(t)|^2=n_e(t)$ being the time-dependent 
populations of the ground and excited states, respectively. 
Normalization of the wave functions implies $n_g(t) + n_e(t) = 1$. 
Using Eq.\ (\ref{eq:2levellin}) to compute the dipole moment 
$d(t)= \langle \psi(t) |\sum_j\hat{\v r}_j | \psi(t) \rangle=d_{eg}2 Re[a_g^*(t)a_e(t)]$ we obtain 
\begin{equation}
d(t)= 2 d_{eg} \sqrt{n_g(t)n_e(t)} \cos(\omega t + \varphi(t)),
\label{eq:d_t_lin}
\end{equation}
where $\varphi(t)$ is a slowly varying (on the scale of $1/\omega$) part of the phase difference of the
coefficients $a_g(t)$ and $a_e(t)$. 
The Rabi dynamics is encoded in the envelope of the dipole moment through the time dependence of $n_g (t)$ and $n_e(t)$.
After the two-level projection the time-dependent Schr\"odinger equation 
$i \partial_t |\psi(t)\rangle = \hat{H} |\psi(t)\rangle $ reduces to the form
\be
i \partial_t  \left( {\begin{array}{c}
 a_g(t)\\
 a_e(t)  \\
 \end{array} } \right)=
\left( {\begin{array}{cc}
 \epsilon_g  &  d_{eg}{\cal E}(t)  \\
d_{eg}{\cal E}(t) & \epsilon_e  \\
 \end{array} } \right) \left( {\begin{array}{c}
 a_g(t)   \\
 a_e(t)
 \end{array} } \right).
\label{eq:proj1}
\ee
The conditions (\ref{eq:cond}) imply the existence of two well-separated timescales: 
one governed by the external frequency $\omega$ and the other determined by the Rabi frequency $\Omega_0$, 
the frequency of the oscillations between the ground and the excited state. This timescale separation allows for 
the use of the rotating wave approximation (RWA) \cite{QMTD} leading to the
following differential equation for $n_e(t)$
(for an outline of the derivation see Appendix A)
\be
\partial^2_t n_e(t)= -\left(\delta^2 + \Omega_0^2\right) n_e(t) + 
\frac{1}{2}\Omega_0^2
\label{eq:n_1_t_t_lin}
\ee
with initial conditions $n_e(0)=0$ and $\partial_t{n}_e(0)=0$. 
Eq.\ (\ref{eq:n_1_t_t_lin}) describes a harmonic oscillator with a restoring 
force which increases with increasing detuning $\delta$ (see potentials for 
different detunigs $\delta$ in Fig.~\ref{fig:rabipot}). Thus, increasing $\delta$ 
results in a squeezing of the harmonic potential leading to a decrease of the amplitude of the oscillations according to
$n_e^{\mathrm{max}}=\Omega_0^2/(\Omega_0^2 + \delta^2)$  and a larger Rabi frequency (see Fig.~\ref{fig:dipolen1Lin}).


In order to investigate the description of Rabi oscillations within TDDFT we 
analyze a one-dimensional (1D) two-electron model system which has the 
advantage that it can be solved exactly \cite{HFCVMTR2011}. For ease of comparison we choose the same model as in \cite{BauerRugg} with the external potential 
\be
V_{ext}= -2/\sqrt{x^2 + 1}
\label{eq:Vext}
\ee
and the electron-electron interaction being of soft-Coulomb type, i.e. $V_{ee}= 1/\sqrt{(x_1 - x_2)^2 + 1}$.
 The eigenfunctions and eigenvalues are obtained by diagonalization of the
 Hamiltonian (\ref{eq:LinHgeneral}) using the {\tt octopus} code \cite{octopus,
 octopus2}. 
The calculations are performed in a box from $-100$ to $100$ bohr with a spacing of $0.2$ bohr. The obtained eigenvalues are $\epsilon_g=-2.238$ Ha and
 $\epsilon_e=-1.705$ Ha, and the static dipole matrix element is $d_{eg}=1.104$.
In order to induce Rabi oscillations, a laser field of the form ${\cal E}(t)={\cal E}_0 \sin(\omega t)$ is turned on at $t=0$, with ${\cal E}_0=0.0125\omega$ 
which ensures conditions (\ref{eq:cond}) are satisfied.
The frequency $\omega$ of the applied field has been chosen to be close to the resonance $\Delta=\epsilon_e-\epsilon_g$.
In Fig.~\ref{fig:dipolen1Lin} the time-dependent dipole moment $d(t)$ and the 
populations $n_g(t)$ and $n_e(t)$ for $\delta= 0.08 \Omega_0$ and 
$\delta=2.2 \Omega_0$ are shown. 
The effect of the detuning manifests in an incomplete population of the 
excited state and a consequent decrease in the amplitude of the envelope of $d(t)$ that is proportional to $\sqrt{n_g n_e}$. 
For small detuning the minima and the maxima of $n_e$ coincide with minima of the envelope, but for larger detuning the dipole
moment only goes to zero for the minima of $n_e$. In 
Fig.~\ref{fig:dipolen1Lin}a, $\delta$ is very small but non-zero leading to the
appearance of a neck at the odd minima of the envelope function which coincide with the minima of $n_g$. The neck increases with increasing $\delta$ and evolves into a maximum 
for Fig.~\ref{fig:dipolen1Lin}b.  Thus, the first minimum of Fig.~\ref{fig:dipolen1Lin}b corresponds to one complete cycle and can be identified 
with the second minimum in \ref{fig:dipolen1Lin}a. We note that looking only at 
the dipole moment is insufficient to discern between resonant and detuned Rabi 
oscillations, only studying the population $n_e$ of the excited state provides access to the complete 
picture. A comparison between the analytic solution of
Eqs.~(\ref{eq:d_t_lin})-(\ref{eq:n_1_t_t_lin}) and the results of the 
time-propagation with the {\tt octopus} code shows perfect agreement, which
confirms that the conditions (\ref{eq:cond}) are fulfilled for the chosen values
of ${\cal E}_0$ and $\omega$. 

\section{Rabi oscillations in the Kohn Sham system}\label{sec:rabiks}

In TDDFT the interacting system is mapped onto a non-interacting KS system which reproduced the correct density 
$\rho(r,t)$ \cite{RG1984,TDDFT2006}.
The time-dependent KS Hamiltonian corresponding to (\ref{eq:LinHgeneral}) is given as
\begin{equation}
\hat{H}_s= \hat{H}_s^0 + \hat{V}_{hxc}^{dyn}(t) + \hat{\v r}\v {\cal E}(t),
\label{eq:NonLinH}
\end{equation}
where the static KS Hamiltonian reads $\hat{H}_s^0= \hat{T} + \hat{V}_{ext} + 
\hat{V}_{hxc}[\rho_0]$. The KS wave functions $\phi_j(\v r)$ are eigenfunctions of $\hat{H}_s^0$ with
eigenvalues $\epsilon_j^{s}$ and the time-dependent density is computed as $\rho(\v r,t)=
\sum_j |\phi_j(\v r,t)|^2$. Here, we are studying a two-electron singlet
system, but within the two-level approximation there is always one unique
orbital  that is getting deoccupied and another that is getting populated,
independent of the number of particles. Thus, the time  evolution only affects
one unique orbital and follows from the KS equation $i \partial_t \phi(\v r, t)
= \hat{H}_s \phi(\v r, t)$  with initial condition $\phi(\v r, t=0)=\phi_g(\v
r)$. The KS equation is non-linear due to the  dependence of the
Hartree-exchange-correlation potential $V_{hxc}$ on the  density which, for two
electrons in a singlet state, is given as  $\rho(\v r,t)= 2 |\phi(\v r,t)|^2$. The time-dependent
dipole moment $d(t)$ is  an explicit functional of the time-dependent density,
i.e.\ $d(t)=\int d^3r \rho(\v r,t)\v r$. The exact KS system reproduces the
exact  many-body density $\rho(\v r,t)$ and, hence, the exact dipole moment
$d(t)$.  However, this need not be true for an approximate functional.
Especially, using  adiabatic approximations has been shown to have a dramatic
effect on the  calculated density during Rabi oscillations \cite{BauerRugg}.

\begin{figure}
\centering
\includegraphics[width=0.47\textwidth]{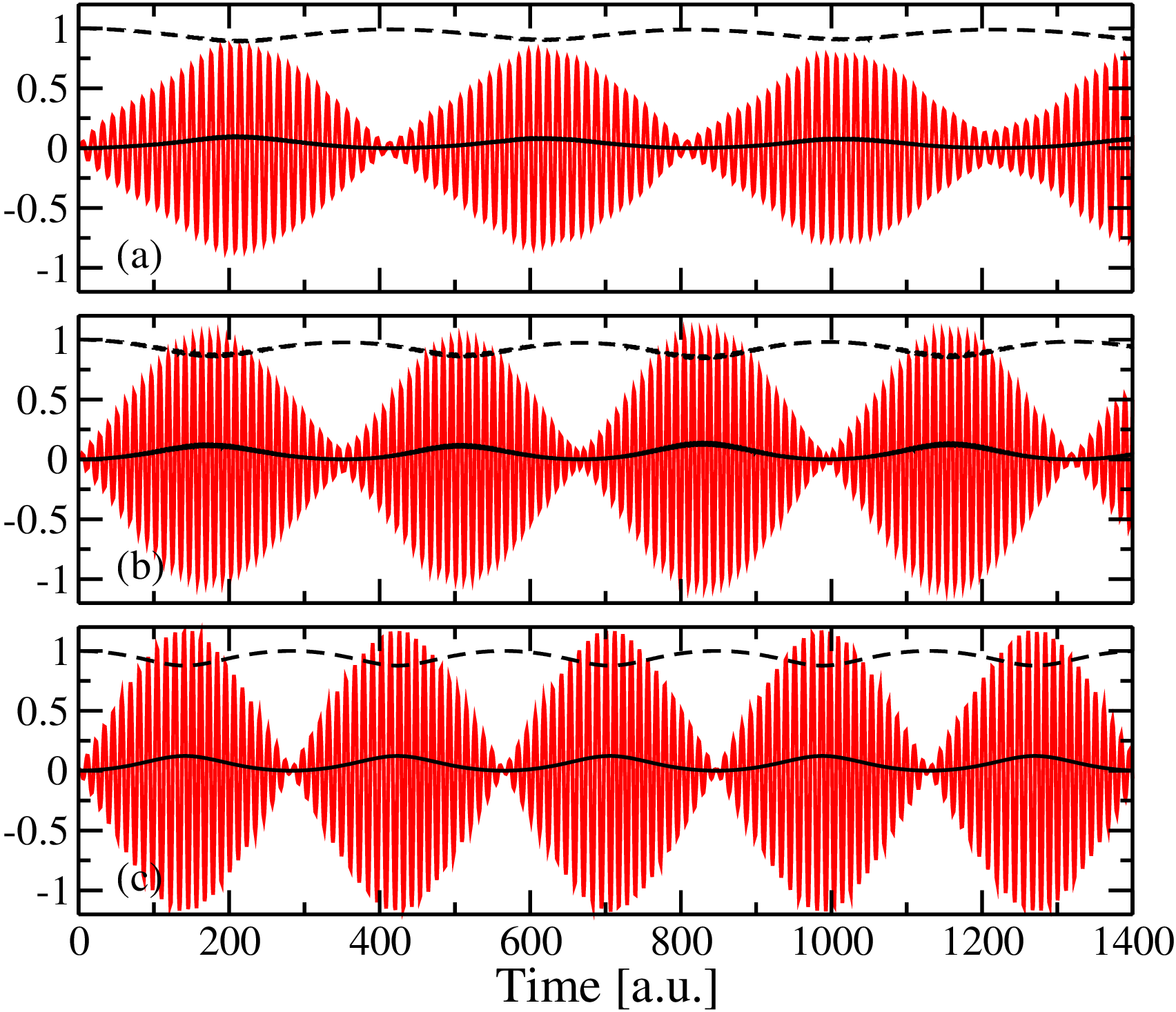}
\caption{Dipole moment (red) and populations $n_e^s$ (solid black line) and 
$n_g^s$ (dashed black line) for ALDA (a) and EXX (b). The results from the theoretical model Eq.\ (\ref{eq:n1_t_t}) are given in (c). The calculations in 
(a) were performed for $\omega^{ALDA}= 0.476$ Ha and the ones in (b) and (c) for $\omega^{EXX}= 0.549$ Ha.}
\label{fig:ExxLDA}
\end{figure}
Here, we employ two different approximations to the one-dimensional xc
potential, the recently derived adiabatic local density approximation (ALDA) \cite{HFCVMTR2011}
and exact exchange (EXX) which for a two-electron singlet is adiabatic and equal to
Hartree-Fock. The resonant frequencies are calculated from linear response in  {\tt octopus} which
yields $\omega^{ALDA}= 0.476$ Ha and $\omega^{EXX}= 0.549$ Ha.  We then apply a
laser field in analogy to the exact calculation with an amplitude of ${\cal
E}_0=0.0125\omega$ using the resonant frequency for each case. Propagating with
the ALDA and EXX results in the dipole moments shown in Fig.~\ref{fig:ExxLDA}a and ~\ref{fig:ExxLDA}b respectively.
Note that despite of the applied laser being in resonance with the system both 
EXX and ALDA approximations show the characteristic  signature of detuned Rabi
oscillations: the population of the excited state is incomplete and the zeros of
the dipole moment coincide with the minima of $n_e^s$ but not with the minima
of $n_g^s$, giving a similar picture as in Fig.~\ref{fig:dipolen1Lin}b. 
As the KS system only has to reproduce the exact density, we do not expect a complete population of the excited KS
 orbital. However, as we can see from Fig.\ \ref{fig:ExxLDA}, the system remains mainly in its ground state and, 
like already pointed out in \cite{BauerRugg}, its density resembles mostly the ground-state density. 
This does not imply that the KS system is not undergoing Rabi oscillations. 
Based on the behavior of the populations we postulate that the oscillations seen in Fig.\ \ref{fig:ExxLDA} 
are in fact detuned Rabi oscillations and not of classical origin as claimed in \cite{BauerRugg}.

If an adiabatic approximation is used the potential at time $t$ is a functional of the density at this time, i.e. $V_{hxc}(t)=V_{hxc}[\rho(t)]$. 
In the following, we show that $V_{hxc}(t)$ indeed introduces a detuning that drives the system out of resonance. 
We again rely on the conditions (\ref{eq:cond}), i.e. describe the KS system as an effective two-level system with  
\be \label{eq:2level}
 \phi(t)(\v r, t) = a_g^s(t) \phi_g (\v r) + a_e^s(t) \phi_e(\v r).
\ee
Projecting the KS Hamiltonian (\ref{eq:NonLinH}) onto the two-level KS space (\ref{eq:2level}) yields the $2\times 2$ matrix
\be
\left( {\begin{array}{cc}
 \epsilon_g^{s} +  \epsilon_g^{xc}(t) &  d_{eg}^{s}{\cal E}(t) +  {\cal F}_{xc}(t) \\
d_{eg}^{s}{\cal E}(t) + {\cal F}_{xc}(t) &  \epsilon_e^{s} + \epsilon_e^{xc}(t) \\
 \end{array} } \right)
\label{eq:proj2}
\ee
with the dipole matrix element being $d_{eg}^{s}=\langle \phi_e|\hat{\v
r} |\phi_g \rangle$. The additional terms,  $\epsilon_{g,e}^{xc}(t)=\langle
\phi_{g,e}|\hat{V}_{hxc}^{dyn}(t)|\phi_{g,e} \rangle$ and ${\cal F}_{xc}(t)=\langle \phi_g|
\hat{V}_{hxc}^{dyn}(t)|\phi_e \rangle$, describe the fictitious time-dependence
that results in a dynamical detuning of the Rabi oscillations. As in the linear
Rabi oscillations, the matrix (\ref{eq:proj2}) determines the coefficients
$a_g^s(t)$ and $a_e^s(t)$ and the equations of motion for the dipole moment
$d^{s}(t)$ and the population $n_e^{s}(t)= |a_e^s(t)|^2$. Compared to
Eq.~(\ref{eq:proj1}) we note that each entry in (\ref{eq:proj2}) contains an
additional term depending on $V_{hxc}^{dyn}$.  In order to investigate the
consequences of this term we again study the external potential (\ref{eq:Vext})
and  use the EXX functional for which a relatively simple analytic expression
for the additional matrix elements can be derived.
The behavior using ALDA is very similar to the one for the EXX approximation (see Fig.~\ref{fig:ExxLDA}a). 
The analysis, however, is more involved due to the functional not being linear in the density. 

\begin{figure}
\begin{center}
\includegraphics[width=0.47\textwidth]{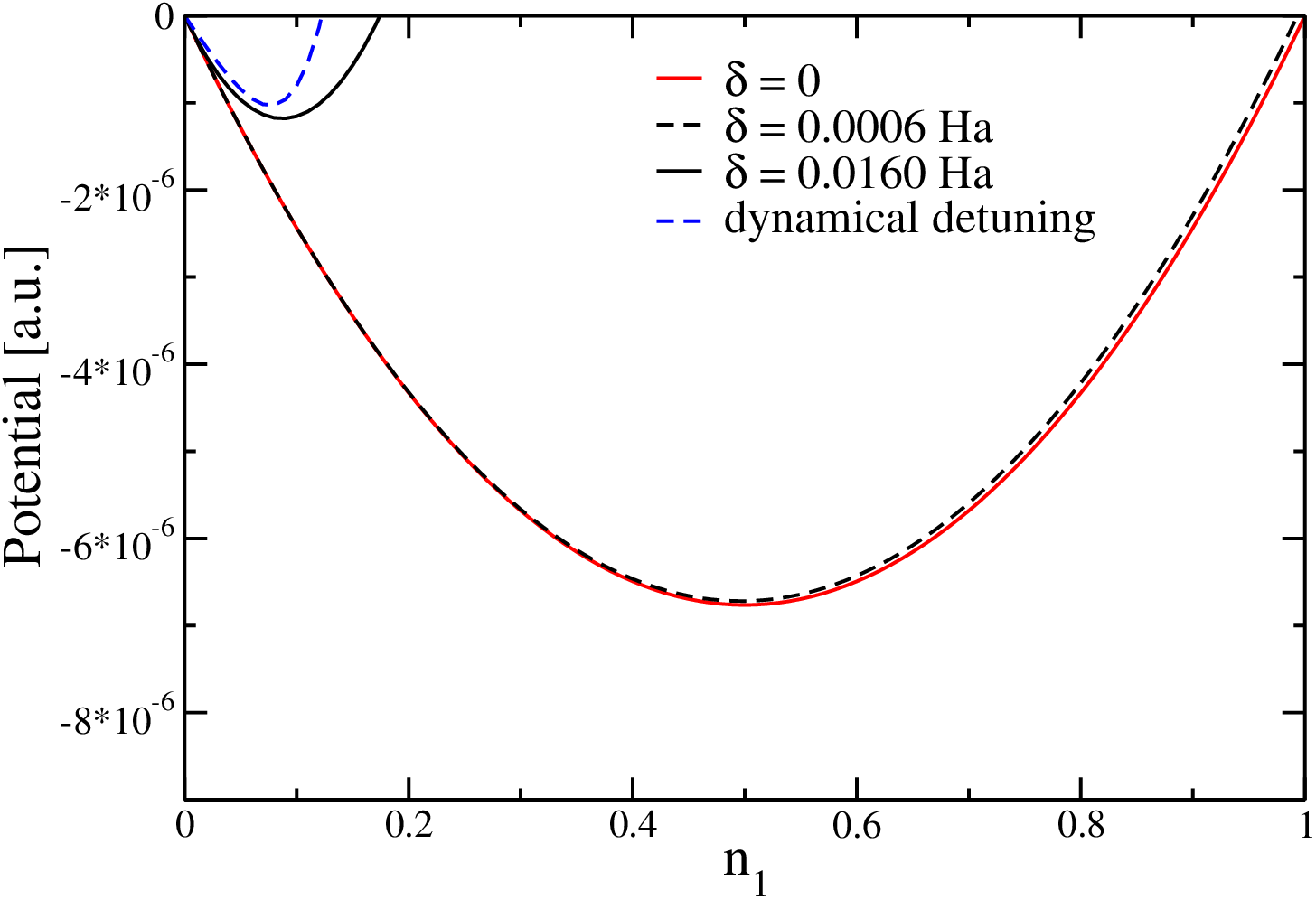}
\end{center}
\caption{\label{fig:rabipot}
Potentials corresponding to the differential equations
(\ref{eq:n_1_t_t_lin}) and (\ref{eq:n1_t_t}).  The dynamical detuning
leads to a quartic potential which has a similar effect to the
potential as a large detuning in the applied frequency (see text for details).}
\end{figure}

For the two-electron singlet case investigated here, the Hartree-exchange-correlation 
potential $V_{hxc}^{EXX}(x,t)$ is equal to half the Hartree potential and, 
hence, given as
\begin{equation}
V_{hxc}^{EXX}(x ,t)= \frac{1}{2}\int d^3x \hat{V}_{ee}( |x- x'|)  (\rho_0(x') + \delta \rho(x',t)).
\label{eq:VHXCEXX}
\end{equation}
Here, the part containing $\rho_0$ determines $V_{hxc}[\rho_0]$ while 
$\delta\rho$ results in the additional $V_{hxc}^{dyn}$. 
We then rewrite the 
contributions to the diagonal terms of Eq.\ (\ref{eq:proj2}) as
\begin{equation}
\epsilon_g^{xc}(t)=\lambda_g n_e^{s}(t), \quad \epsilon_e^{xc}(t)=\lambda_e n_e^{s}(t)
\end{equation}
where, in EXX, the coefficient $\lambda$ reads
\begin{equation}
\lambda_{g,e} \!= \! \int\!\!\!\!\int\!\! dx\: dx' 
\left( |\phi_e(x')|^2 - |\phi_g(x')|^2\right) 
\hat{V}_{ee}( |x - x'|) |\phi_{g,e}(x)|^2.
\label{eq:lambda}
\end{equation}
For the off-diagonal contributions we recall that 
$d^{s}(t)= 2d_{eg}^{s}Re[a_g^s(t)^*a_e^s(t)]$, as in the exact case, and rewrite 
the contribution of ${V}_{hxc}^{dyn}(t)$ to the off-diagonal terms as a 
coefficient $g$ multiplied by the time-dependent dipole moment
\begin{equation}
 {\cal F}_{xc}(t)= g \frac{d^{s}(t)}{d_{eg}^s}.
\label{eq:Fxc}
\end{equation}
Here, g is given as
\begin{equation}
g= \int\!\!\!\!\int dx dx'  \phi_e(x')\phi_g(x')  \hat{V}_{ee}( |x - x'|) 
\phi_g(x)\phi_e(x).
\label{eq:g}
\end{equation}
The coefficient $g$ also enters when one calculates the resonant frequencies in linear response. Within the two-level approximation the resonant 
frequency is given as $\omega_0^{s}=\sqrt{\Delta_s (\Delta_s + 2g)}$ (see Appendix B) which  yields $\omega_0^{EXX}= 0.532$~Ha. 
The deviation from $\omega^{EXX}$ calculated from time propagation of Hamiltonian (\ref{eq:NonLinH}) in {\tt octopus} is of the order of $3\%$, 
coinciding with the deviation of our system from a true two-level system which we estimate from $\left(1-(n_g^s(t)+n_e^s(t))\right)$.
Using again the RWA we obtain, to leading order in $\lambda/\omega_0$ and $g/\omega_0$, the following differential equation for the level 
population $n_e^s(t)$ (see Appendix B for the derivation)  
\be
\partial^2_t n_e^s(t) =  - \left(\frac{\gamma^2}{2}n_e^s(t)^2 +
\Omega_{s}^2\right) n_e^s(t) +\frac{1}{2}\Omega_{s}^2
\label{eq:n1_t_t}
\ee
with $\Omega_{s}=d_{eg}^{s}{\cal E}_0$ and $\gamma=\lambda - 2 g$. Neglecting
higher order terms leads to an error of about 10$\%$. Within these error bars we
can safely use Eq.\ (\ref{eq:n1_t_t}) because it contains all the relevant 
physics. Unlike
Eq.~(\ref{eq:n_1_t_t_lin}) which represents a harmonic oscillator,
Eq.~(\ref{eq:n1_t_t}) corresponds to an anharmonic quartic oscillator and its
solution can be written in terms of Jacobi elliptic functions \cite{AQO}.
Equivalently, Eq.~(\ref{eq:n1_t_t}) can be integrated numerically.  Comparing
Eqs.\ (\ref{eq:n1_t_t}) and (\ref{eq:n_1_t_t_lin}) we conclude that the
adiabatic approximation introduces a time-dependent detuning proportional  to
$\frac{1}{\sqrt{2}} \gamma n_e^{s}(t)$. The anharmonic oscillator no longer represents a
parabola  but the dynamical detuning also results in an increase of the
restoring force. This can be clearly seen in Fig.~\ref{fig:rabipot} where we
plot the potential corresponding to the restoring force in Eqs.\
(\ref{eq:n_1_t_t_lin}) and (\ref{eq:n1_t_t}) for the detunings of Fig.\
\ref{fig:dipolen1Lin} and the EXX calculation( Fig.\ \ref{fig:ExxLDA}c).

Using the same 1D model system Eq.~(\ref{eq:Vext}) as before we apply  a field of
amplitude ${\cal E}_0=0.0125\omega^{EXX}$ and calculate the dipole moment
$d^s(t)$ and population $n_e^s(t)$ from Eq.~(\ref{eq:n1_t_t}).  For this system
we obtain for the bare KS eigenvalues $\epsilon_g^s=-0.750$~Ha and
$\epsilon_e^s=-0.257$~Ha which yields $\Delta_s=0.494$~Ha.  For the various
matrix elements we obtain $d_{eg}^{s}=0.897$,  $g=0.071$, $\lambda= -0.125$, and
$\gamma=-0.268$. The results are shown in Fig.\ \ref{fig:ExxLDA}c in comparison
to the numerically exact time-propagation  in {\tt octopus} (Fig.\ \ref{fig:ExxLDA}b).  The
discrepancy between the numerical propagation and the analytical results is
mainly due to the fact that we kept only the leading orders in $\lambda/\omega_0$ and
$g/\omega_0$ in the derivation of Eq.~(\ref{eq:n1_t_t}). However, our simple
model clearly captures the effect of the dynamical detuning present in all
adiabatic functionals.  In the physical system the density changes dramatically
during the transition and, in order to be able to describe resonant Rabi
oscillations, a functional has to keep track of these changes over time, i.e.\
it needs to have memory.  For any adiabatic functional the potential will change
due to the changing density and the system is driven out  of resonance. We
emphasize that this effect is not limited to TDDFT but is generic for all
mean-field theories,  e.g.\ HF or all hybrids, when the effective potential
depends instantaneously on the state of the system.

\section{Conclusion}\label{sec:concl}
In conclusion, we demonstrate that the use of adiabatic approximations leads to
a dynamical detuning  in the description of Rabi oscillations. Only the
inclusion of an appropriate memory dependence can correct the fictitious
time-dependence of the resonant frequency. In particular, by clearly identifying
the reason behind the failure of adiabatic approximations, we are able to prove
the correspronding conjecture recently made in \cite{BauerRugg}. Our results
constitute a very stringent test for the development of new xc functionals
beyond the linear regime as all (adiabatic) functionals available till now fail
to reproduce Rabi dynamics. Adiabatic functionals will fail similarly in the
description of all processes involving a change in the population of states. 
Work along the lines of deriving a new memory-dependent functional is ongoing.
The  description of photo-induced processes in chemistry, physics, and
biology and the new field of attosecond electron dynamics  and high-intense
lasers all demand fundamental functional developments going beyond the adiabatic
approximation.

\acknowledgments

We acknowledge support by  MICINN (FIS2010-21282-C02-01), ACI-promociona
(ACI2009-1036), Grupos Consolidados UPV/EHU del Gobierno Vasco (IT-319-07), the
European Community through e-I3 ETSF project (Contract No. 211956), and the
European Research Council Advanced Grant DYNamo (ERC-2010-AdG -Proposal No.
267374). JIF acknowledges support from an FPI-fellowship (FIS2007-65702-C02-01).

\appendix

\section{Equation of motion for $n_e(t)$ in the interacting many-body problem}

We provide here a derivation of the equation of motion, Eq.~(\ref{eq:n_1_t_t_lin}), for the population $n_{e}(t)$ of the excited state. 

As a first step we formulate a system of equations for physical observables, the dipole moment  $d(t)=2d_{eg}Re[a_g^*a_e]$, the ``transition current'' $J(t)=2d_{eg}Im[a_g^*a_e]$, and 
the populations $n_{g,e}(t)=|a_{g,e}|^2$.
Using the Schr\"odinger equation Eq.~(\ref{eq:proj1}) for $a_{g,e}(t)$ and the above definitions of $d(t)$, $J(t)$ and $n_{g,e}(t)$  one can derive the following coupled differential equations for these quantities,
\bea
\label{eq:d_t}
\partial_t d(t) &=& \Delta J(t),\\
\label{eq:q_t}
\partial_t J(t) &=& -\Delta d(t) - 2 (n_g - n_e)|d_{eg}|^2{\cal E}(t),\\
\label{eq:pop}
\partial_t n_e(t) &=& -J(t) {\cal E}(t),
\eea
where $\Delta=\epsilon_e-\epsilon_g$ with $\epsilon_{g,e}=\langle
\psi_{g,e}|\hat{H}(t)|\psi_{g,e} \rangle$ and $d_{eg}{\cal E}(t)=\langle
\psi_{e}|\hat{H}(t)|\psi_{g} \rangle$. Equations (\ref{eq:d_t}), (\ref{eq:q_t}), and (\ref{eq:pop}) 
correspond to the two-level version of the continuity equation, the equation of motion for the transition current, 
and the energy balance equation, respectively. These equation have to be supplemented with the normalization condition $n_g(t)+n_e(t)=1$, and the initial conditions $d(0)=J(0)=n_{e}(0)=0$.  

By combining Eqs.~(\ref{eq:d_t}) and (\ref{eq:q_t}) one obtains an equation of motion for the dipole moment $d(t)$
\be
\partial^2_t d(t)= -\Delta^2 d(t) -2 (1 - 2n_e)\Delta |d_{eg}|^2{\cal E}(t).
\label{eq:d_t_t_lin}
\ee
The next step in the derivation is to simplify this equation using RWA. Namely, we separate ``fast'' and ``slow'' time scales by writing the dipole moment in the following form
\be
d(t)= d_{eg}[b_1(t) \cos(\omega t)+ b_2(t) \sin(\omega t)]
\label{eq:ansatz_d}
\ee
where the coefficients $b_1(t)$ and $b_2(t)$ are assumed to be slowly varying ($\partial_t{b}/b\ll\omega=\Delta+\delta$). 
Inserting the ansatz of Eq.~(\ref{eq:ansatz_d}) into Eq.~(\ref{eq:d_t_t_lin}) and making use of the condition
$\delta,\Omega_0<<\Delta$ we arrive at the system of first order differential equations for slow variables
\bea
\label{eq:b_2_t}
\partial_t b_2(t) - \delta b_1(t) &=& 0\\
\label{eq:b_1_t}
\partial_t b_1(t)+\delta b_2(t) - \Omega_0(1 - 2n_e) &=&0
\eea
From Eqs.~(\ref{eq:pop}) and (\ref{eq:d_t}) we obtain the equation of motion for the populations
\be
\partial_t n_e = \frac{1}{2}\Omega_0b_1(t).
\label{eq:nu_t}
\ee
where we again kept only leading in $\omega$ terms and, as usual in RWA, neglected irrelevant terms 
oscillating with $2\omega$. 
Combining this equation with Eq.~(\ref{eq:b_2_t}) results in 
\be
\partial_t\left[b_2(t)-\frac{2\delta}{\Omega_0}n_e(t)\right]=0
\ee
which, together with the initial conditions $n_e(0)=0$ and $\partial_t d(0)=0$, yields 
\be
b_2(t)= \frac{2\delta}{\Omega_0}n_e(t)
\label{eq:b_2}
\ee
and, due to Eq.~(\ref{eq:b_2_t}), 
\bea
b_1(t)&=&\frac{2}{\Omega_0}\partial_t n_e(t)
\label{eq:b_1}
\eea

Finally, by inserting Eq.~(\ref{eq:b_2}) and Eq.~(\ref{eq:b_1}) into
 Eq.~(\ref{eq:b_1_t}) we arrive at the differential equation Eq.~(\ref{eq:n_1_t_t_lin}) for the population $n_e(t)$ of 
the excited state in the many-body interacting system.

\section{Equation of motion for $n_e^s(t)$ in the KS system}

Following the same scheme as in appendix A we do now derive Eq.~(\ref{eq:n1_t_t}) describing dynamics of the 
KS population $n_{e}^s(t)$. The equations of motion for the KS quantities, $d^s(t)=2d_{eg}^sRe[(a_g^s)^*a_e^s]$, 
$J^s(t)=2d_{eg}^sIm[(a_g^s)^*a_e^s]$, and $n_{g,e}^s(t)=|a_{g,e}^s|^2$ are similar to Eqs.~(\ref{eq:d_t})-(\ref{eq:pop}) and
can be obtained with the following replacements:$\Delta \rightarrow \Delta(t)=  \Delta_s +
\epsilon_e^{xc}(t)-\epsilon_g^{xc}(t)$, and
$d_{eg}{\cal E}(t) \rightarrow d_{eg}^{s}{\cal E}(t) + {\cal F}_{xc}(t)$. One
can again derive an equation of motion for the dipole moment (in analogy to 
Eq.~(\ref{eq:d_t_t_lin})) which, due to the additional time dependence in
$\Delta(t)$, aquires an additional term and reads
\be
\begin{split}
\partial^2_t {d^s(t)} = & -\Delta^2(t){d^s(t)}
+\frac{\partial_t\Delta(t)}{\Delta(t)}{\partial_t d^s(t)}\\ 
& -2 (n_g^s - n_e^s)\Delta(t)d_{eg}(d_{eg}^s{\cal E}(t) +{\cal F}_{xc}(t)),
\end{split}
\label{eq:d_t_t}
\ee
where ${\cal F}_{xc}(t)$ is given by Eq.~(\ref{eq:Fxc}).
For the two-electron singlet case and EXX approximation, 
the time-dependent contribution to the Hartree-exchange-correlation potential 
is proportional to $\delta\rho(x,t)$. As $\delta \rho(x,t) = 2n_e^s(t)
(|\phi_e(x)|^2 - |\phi_g(x)|^2) + 2d^s(t)\phi_g(x)\phi_e(x)$ we obtain
\be
\Delta(t) = \Delta_s +\lambda n_e^s(t)
\ee
with $\lambda=\lambda_e-\lambda_g$ and $\lambda_{g,e}$ defined in
Eq.~(\ref{eq:lambda}). 

The resonant frequency $\omega_0^{EXX}$ is now not given by the KS energy
difference $\Delta_s$ but needs to be calculated from the linear response. 
In the linear regime Eq.~(\ref{eq:d_t_t}) reduces to the following form
\be
\partial^2_t {d^s(t)} = -\Delta_s\left((\Delta_s+2g)d^s(t)-2\Delta_s|d_{eg}^s|^2{\cal E}(t)\right),
\label{eq:d_t_t_EXX}
\ee
From the first term in the right hand side in Eq.~(\ref{eq:d_t_t_EXX}) we identify the linear response resonant 
frequency as
\bea
\omega^{EXX}_0=\sqrt{(\Delta_s+2g)\Delta_s}.
\label{eq:0mega_0_EXX}
\eea

We can now apply the same procedure to Eq.~(\ref{eq:d_t_t}) as in the 
interacting case to derive an equation of motion for the occupation $n_e^s(t)$. 
Below, as well as in the main text, for the KS system we consider only the case of a resonant excitation, 
$\omega=\omega^{EXX}_0$, i.~e. $\delta=0$.
Employing the ansatz of Eq.~(\ref{eq:ansatz_d}) for the KS dipole moment $d^s(t)$, in analogy to Eqs.~(\ref{eq:b_2_t}) and (\ref{eq:b_1_t}) we obtain
\bea
\label{eq:b_2_t_EXX}
\partial_t b_2^s(t)+\left(\lambda- 2g\right)n_e^s(t)b_1^s(t)&=& 0\\
\nonumber
\partial_t b_1^s(t) - \left(\lambda- 2g\right)n_e^s(t)b_2^s(t)&&\\
 -\Omega_s(1-2n_e^s(t))&=& 0
\label{eq:b_1_t_EXX}
\eea

with $\Omega_s=d_{eg}^s{\cal E}_0$. From the EXX analog of Eq.~(\ref{eq:pop}) (the energy balance equation) we find that to the leading order in RWA the equation of motion for the occupation has exactly the same form as Eq.~(\ref{eq:nu_t}), i.~e.,
\bea
\partial_t n_e^s(t) = \frac{1}{2}\Omega_s b_1^s(t).
\eea

Therefore, with the initial conditions $n_e^s(0)=0$ and $\partial_t n_e^s(0)=0$ we get for the two coefficients
\bea
\label{eq:b_1_EXX}
b_1^s(t)&=& \frac{2}{\Omega_s}\partial_t n_e^s(t),\\
\label{eq:b_2_EXX}
b_2^s(t)&=&-\frac{\lambda-2g}{\Omega_s} (n_e^s(t))^2.
\eea
Finally, by inserting Eq.~(\ref{eq:b_1_EXX}) and Eq.~(\ref{eq:b_2_EXX}) into 
Eq.~(\ref{eq:b_1_t_EXX}) we arrive at the differential equation Eq.~(\ref{eq:n1_t_t}) for the population $n_e^s(t)$ 
of the KS excited state.

\end{document}